\documentclass[aps,prb,reprint,twocolumn,nofootinbib,longbibliography,superscriptaddress,draft=false]{revtex4-1}

\usepackage{amsmath, amsfonts, pifont, bm, bbm, braket, mathtools, comment}
\usepackage[inline]{enumitem}
\usepackage{hyperref}
\hypersetup{colorlinks=true,
		    linkcolor=blue,
		    citecolor=blue,
		    allcolors=blue}
\usepackage{natbib}
\newcommand*{\citen}[1]{%
  \begingroup
    \romannumeral-`\x 
    \setcitestyle{numbers}
    [\cite{#1}]
  \endgroup   
}
\usepackage{graphicx}

\newcommand{\pmat}[1]{\begin{pmatrix} #1 \end{pmatrix}} 
\renewcommand{\d}{\mathrm{d}}
\newcommand{\id}{\mathbbm{1}}

\DeclareMathOperator{\sgn}{\mathrm{sgn}}

\newcommand{\rucl}{$\alpha$-RuCl$_3$}
\newcommand{\cmark}{\text{\ding{51}}}
\newcommand{\xmark}{\text{\ding{55}}}

\newcommand{\Z}{\mathbb{Z}}
\renewcommand{\H}{\mathcal{H}}
\newcommand{\F}{\mathcal{F}}
\newcommand{\D}{\mathcal{D}}

\renewcommand{\S}{\bm{S}}

\newcommand{\Wf}{\Psi}

\newcommand{\vp}{\varphi}
\newcommand{\ve}{\varepsilon}

\newcommand{\s}{\sigma}

\newcommand{\x}{\bm{x}}

\newcommand{\p}{\bm{p}}
\newcommand{\q}{\bm{q}}
\renewcommand{\a}{\bm{\hat{a}}}
\renewcommand{\b}{\bm{\hat{b}}}
\renewcommand{\c}{\bm{\hat{c}^*}}
\renewcommand{\t}{\bm{\tau}}
\newcommand{\m}{\bm{m}}
\newcommand{\h}{\bm{h}}
\newcommand{\hh}{\bm{\hat{h}}}

\newcommand{\gh}{\bm{\hat{g}}}

\begin{document}

\title{Testing Ising Topological Order in $\mathbf{\alpha}$-RuCl$_\mathbf{3}$ Under In-Plane Magnetic Fields}

\author{Jacob S. Gordon}
\affiliation{Department of Physics, University of Toronto, Ontario M5S 1A7, Canada}
\author{Hae-Young Kee}
\email{hykee@physics.utoronto.ca}
\affiliation{Department of Physics, University of Toronto, Ontario M5S 1A7, Canada}
\affiliation{Canadian Institute for Advanced Research, CIFAR Program in Quantum Materials, Toronto, ON M5G 1M1, Canada}

\date{\today}

\begin{abstract}


  Material realization of the non-Abelian Kitaev spin liquid phase - an example of Ising topological order (ITO) - has been the subject of intense research in recent years.
  The $4d$ honeycomb Mott insulator \rucl\ has emerged as a leading candidate, as it enters a field-induced magnetically disordered state where a half-integer quantized thermal Hall conductivity $\kappa_{xy}$ was reported.
  Further, a recent report of a sign change in the quantized $\kappa_{xy}$ across a certain crystallographic direction is strong evidence for a topological phase transition between two ITOs with opposite Chern numbers.
  Although this is a fascinating result, independent verification remains elusive, and one may ask if there is a thermodynamic quantity sensitive to the phase transition.
  Here we propose that the magnetotropic coefficient $k$ under in-plane magnetic fields would serve such a purpose.
  We report a singular feature in $k$ that indicates a topological phase transition across the $\b$-axis where ITO is prohibited by a $C_2$ symmetry.
  If the transition in \rucl\ is indeed a direct transition between ITOs, then this feature in $k$ should be observable.
  
\end{abstract}

\maketitle

\noindent{\it Introduction} -- The Kitaev honeycomb model~\cite{kitaev2006anyons} is an important example of an exactly solvable spin-$\tfrac{1}{2}$ system displaying topological order.
Spins in the Kitaev model fractionalize into itinerant Majorana fermions (MFs) on a static $\Z_2$ gauge field, which have a Dirac dispersion when the bond-dependent coupling constants are nearly isotropic.
A perturbative magnetic field retains the exact solvability of the model, and serves to gap out the MFs and impart them with a non-trivial Chern number.
The resulting gapped phase exhibits \emph{Ising topological order} (ITO), which hosts non-Abelian anyon excitations and a chiral edge MF which manifests as a half-integer quantized thermal Hall conductivity $\kappa_{xy}$.

Remarkable properties of this model have incited a flurry of research into its material realization.
One mechanism to generate the bond-dependent Kitaev interactions between $j_{\mathrm{eff}} = \tfrac{1}{2}$ states was proposed by \citet{jk2009prl}, and found relevance in honeycomb $d^5$ Mott insulators with strong spin-orbit coupling.
Later, Rau et. al. derived the generic nearest-neighbour interactions in such a system, and found additional off-diagonal symmetric interactions dubbed $\Gamma$ and $\Gamma^{\prime}$~\cite{rau2014prl,rau2014trigonal}.
In parallel, promising material candidates such as the iridates A$_2$IrO$_3$~\cite{jk2009prl,cjk2010prl,singh2012relevance,modic2014realization} (A = Li, Na) and \rucl~\cite{plumb2014prb,sandilands2015continuum,HSKim2015prb,banerjee2016proximate,sandilands2016excitations} were identified, and shown to exhibit magnetic ordering at low temperatures~\cite{fletcher1967magnetic,choi2012prl,cjk2014zigzag,sears2015prb,johnson2015monoclinic,cao2016structure,HSKim2016structure,janssen2017model}.
Shortly after it was found that an applied magnetic field can destroy the underlying magnetic order~\cite{liu2018dirac,lampenkelley2018induced,liu2018dirac,baek2017evidence,wolter2017induced,zheng2017gapless,janvsa2018observation}, and excitement was built around the prospect of an intermediate spin-liquid phase.

\begin{figure}[tp]
\includegraphics[width=\columnwidth]{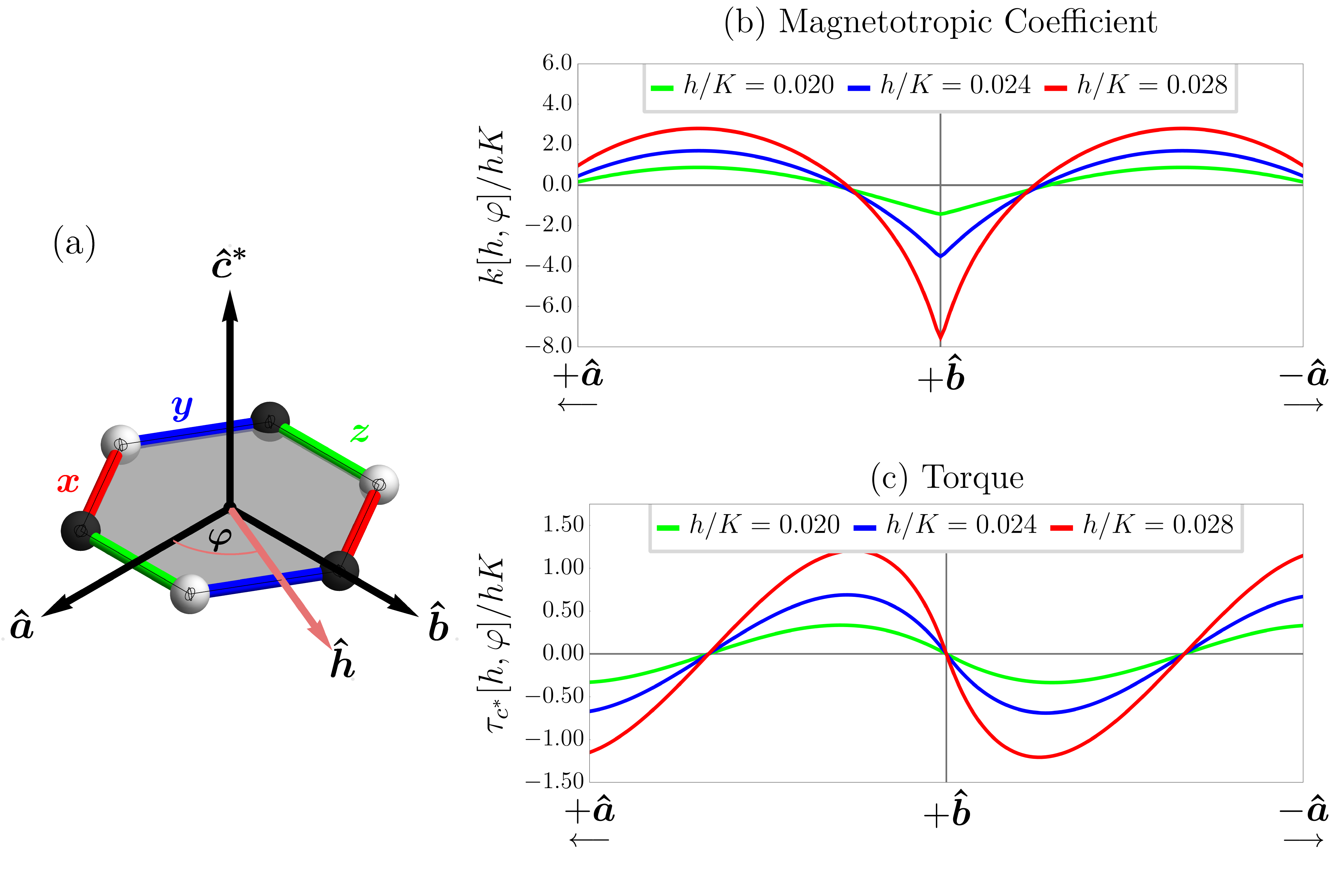}
\caption{
  (a) In-plane magnetic field angle $\vp$ with respect to the honeycomb $\a$- and $\b$-axis, with the bond types indicated.
  (b) The magnetotropic coefficient $k$ and (c) torque as a function of $\vp$ at different field strengths.
  While the torque is not clearly singular, the magnetotropic coefficient $k$ is sensitive to the transition and develops a cusp across the $C_2$-preserving $\b$-axis.
  See main text for details.
}
\label{fig1}
\end{figure}

Most recently, a half-integer quantized thermal Hall conductivity was reported by \citet{kasahara2018thermal} in \rucl\ under a small range of magnetic fields ($\sim 7$-$10$ T) where the magnetic order is suppressed - a smoking-gun signature of a chiral edge MF associated with the bulk ITO.
A follow-up study~\cite{yokoi2020anomalous} reported a varied sign structure of $\kappa_{xy}$ when the field is rotated in the honeycomb plane, indicating a topological phase transition between ITOs with opposite Chern number.
While this is an exciting development, the thermal transport measurements have been notoriously hard to reproduce, and other interesting probes~\cite{aasen2020electrical} may be out of reach for some time.
One may wonder, then, if there is a thermodynamic quantity that is sensitive to these transitions.
Here we propose that the magnetotropic coefficient~\cite{modic2018resonant,modic2019scale} $k$ as a function of the in-plane magnetic field orientation would serve this purpose.
If the transition in \rucl\ between ITOs with different Chern numbers is indeed present in a range of intermediate in-plane magnetic field strengths, then a singular feature in $k$ should be observable as the field angle $\vp$ sweeps across the $\b$ axis as shown in Fig.~\ref{fig1}(a). \newline

\noindent{\it Ising Topological Order} -- A partial characterization of ITO is through its anyonic excitations; one is a MF $\psi$ and the other is a vortex $\sigma$.
These excitations obey the following fusion rules,
\begin{equation}\label{eq:fusion}
  \psi\otimes\psi = \id \qquad
  \sigma\otimes\sigma = \id\oplus\psi \qquad
  \psi\otimes\sigma = \sigma,
\end{equation}
as well as certain associative and braiding relations.
In particular, $\sigma$ is the non-Abelian anyon as their fusion has more than one outcome.
ITO can emerge in a variety of systems, including the Kitaev honeycomb model when a small magnetic field is applied to the gapless phase.
The conical dispersion of MFs is gapped by the applied field, and they acquire a non-zero Chern number $\nu$ which governs the edge thermal transport.
If $\nu$ is odd, then the vortex excitations of the $\Z_2$ gauge field carry an unpaired MF which is responsible for their non-Abelian statistics and ultimate realization of ITO~\cite{kitaev2006anyons}.
No unpaired MF accompanies vortices for even $\nu$, and the resulting phase is an Abelian topological order. \par

Symmetry places important constraints on ITO, which can be easily understood through a precise analogy.
ITO can be regarded as a spinless $p_x + i p_y$ superconductor after gauging the $\Z_2$ fermionic parity of the Bogoliubov quasiparticles.
Time-reversal symmetry (TRS, $\Theta$) is clearly broken, and under $\Theta$ the gap function maps to $p_x - i p_y$ while the chirality of the MF edge mode is reversed.
If the system possesses a $C_2$ or mirror symmetry, say, a $C_2$ about $\bm{\hat{y}}$, then the gap function maps to $-p_x + i p_y$ and again the chirality of the edge mode is reversed.
When both TRS and such a two-fold symmetry are present, ITO is forbidden unless the latter is spontaneously broken~\cite{zou2020QCD3}.
This may occur if TRS is broken by an external magnetic whose direction is chosen such that the two-fold symmetry is maintained.
If this field is swept across the symmetric direction, then, barring any spontaneous symmetry breaking, there are two possibilities.
One is a direct transition between ITOs with opposite Chern numbers (i.e. time-reversal partners), accompanied by a sign change in $\kappa_{xy}$.
Another is an intermediate phase surrounding the symmetric direction with an even Chern number.
In either case there must be at least one phase transition around these special directions, and are therefore prime locations to seek them out. \par

\rucl\ and the iridate candidates consist of quasi-2D honeycomb layers, with a space group symmetry of $C2/m$ or $R\overline{3}$~\cite{sears2015prb,cao2016structure,fletcher1967magnetic,HSKim2016structure,park2016emergence,janssen2020magnon}.
Both groups contain a $C_2(\b)$ rotation about the crystallographic $\b$ direction $[\overline{1}10]$ in addition to translations $T_{1,2}$.
Under $C_2(\b) = e^{-i\pi\b\cdot\S}$ the pseudospin-$\tfrac{1}{2}$ operators transform as
\begin{align}
  \begin{split}
    C_2(\b) S^x_j C_2^{\dag}(\b) &= -S^y_{C_2(\b) j} \\
    C_2(\b) S^y_j C_2^{\dag}(\b) &= -S^x_{C_2(\b) j} \\
    C_2(\b) S^z_j C_2^{\dag}(\b) &= -S^z_{C_2(\b) j},
  \end{split}
\end{align}
from which we can see that a Zeeman term $-\h\cdot\S$ preserves $C_2(\b)$ for $\hh \parallel \b$.
The $R\overline{3}$ space group contains two additional $C_2$ operations related to $C_2(\b)$ through $C_3(\c)$.
For fields in the $\a\b$ plane shown in Fig.~\ref{fig1}(a) and an $R\overline{3}$ space group, a $C_2$ symmetry is preserved for $\vp = \tfrac{\pi}{6}(2n+1)$, while it is preserved in $C2/m$ only for $\vp = \pm \tfrac{\pi}{2}$ ($\hh = \pm \b$).
A summary of these symmetries can be found in Table~\ref{table1} for an $R\overline{3}$ space group. \par

\begin{table}[h!]
  \centering
  \begin{tabular}{||c | c | c | c | c ||}
    \hline
    \hspace{0.5cm} $\bm{\vp}$ \hspace{0.5cm} & \hspace{0.5cm} $\bm{\Theta}$ \hspace{0.5cm} & \hspace{0.5cm} $\bm{C_2}$ \hspace{0.5cm} & \hspace{0.5cm} {\bf ITO} \hspace{0.5cm} \\
    \hline\hline
    $\hspace{5mm}\tfrac{\pi}{6}(2n+1)$ \hspace{0.5cm} & $\xmark$ & $\cmark$ & $\pmb{\xmark}$ \\
    \hline
    $\mathrm{else}$ & $\xmark$ & $\xmark$ & $\pmb{\cmark}$ \\
    \hline
  \end{tabular}
  \caption{
    An external in-plane magnetic field $\hh = \cos(\vp)\a + \sin(\vp)\b$ breaks TRS $\Theta$, but maintains a $C_2$ symmetry for special directions.
    A $C_3$ symmetric $R\overline{3}$ space group is assumed so that there are three distinct $C_2$ operations.
  }
  \label{table1}  
\end{table}

Due to the protected mirror symmetry along the $\b$-axis, there is no ITO when the field is along this direction, and a direct transition between two ITO with opposite $\nu$ is possible as the field sweeps across the $\b$-axis.
While the sign change of the quantized $\kappa_{xy}$ reported in \rucl\ provides evidence of such a transition, one may ask if there are other experimental quantities that may corroborate this finding.
Below we show the magnetotropic coefficient, the second derivative of the free energy with respect to the in-plane angle $\vp$, is a way to independently verify this transition.
Assuming the reported transition in \rucl\ across the $\b$ axis in the range of $7$-$10$ T is directly between ITOs with 
opposite $\nu$, there should be an associated cusp singularity in $k$. \newline

\noindent{\it Magnetotropic Coefficient of \rucl} -- We begin with a brief review of the magnetotropic coefficient $k$, then apply this idea to the exactly solvable Kitaev model used in~\citen{yokoi2020anomalous} to explain the sign structure of $\kappa_{xy}$.
A cusp singularity in $k$ across the $\b$ axis will be demonstrated, as seen in Fig.~\ref{fig1}(b), and traced to the behaviour of the bulk fermion gap. \par

\begin{figure*}[tp]
\includegraphics[width=\linewidth]{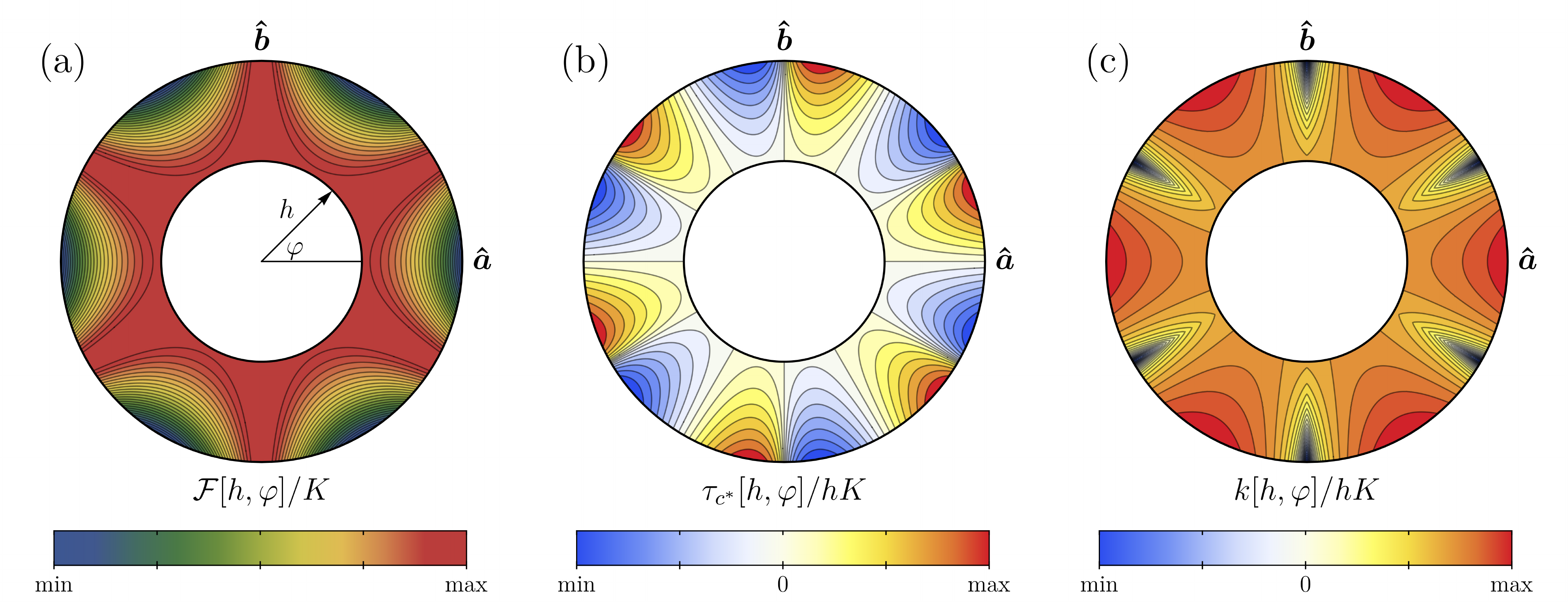}
\caption{
  Thermodynamic properties calculated with the free energy, where the radius represents the magnetic field strength.
  (a) Free energy $\F$ displaying maxima along three $C_2$-symmetric directions for an $R\overline{3}$ space group.
  (b) The $\c$ component of the torque which changes sign across the crystallographic axes as well as the $C_2$-symmetric lines.
  (c) Magnetotropic coefficient $k$ shows most clearly the transitions between ITOs with opposite Chern number.
  The midpoints of the ITO phases display a broad, positive peak in $k$ while the ITO transition points are marked by a sharp, negative cusp indicating an instability with respect to a rotation of the field.
}
\label{fig2}
\end{figure*}

Familiar thermodynamic quantities such as the magnetization $\m$ and torque $\t$, can be defined as the first derivative of the free energy $\F$ with respect to the magnetic field strength and orientation, respectively, and are widely used as indicators of phase transitions.
In particular, derivatives of $\F$ with respect to the field strength and angle yield the magnetization along the field direction and the torque perpendicular to the plane of rotation, respectively.
However, susceptibilities, involving second derivatives of $\F$, can offer more fundamental insights into the nature of phase transitions.
In the same spirit as the magnetic susceptibility, the magnetotropic coefficient $k$ is the second derivative of $\F$ with respect to the field angle - or equivalently the first derivative of the perpendicular component of the torque.
Since torque must vanish along the crystallographic axes, the magnetotropic coefficient has a distinct advantage in studying transitions across these directions.
Resonant torque magnetometry~\cite{modic2018resonant} was developed as a tool to directly measure $k$ through a shift in the natural frequency of a cantilever, and has allowed for unprecedented accuracy in measurements of magnetic anisotropy.
This technique was recently applied to \rucl\ to determine the phase boundary between long-range zig-zag order and the intermediate non-magnetic phase in the $\a\c$ plane~\cite{modic2018resonant,modic2019scale}.
While the torque only displays a deviation from the linear response regime of $\propto h^2$, the magnetotropic coefficient signals this transition with a sharp jump, demonstrating its sensitivity to the phase boundary. \par

Since we expect the response of the Kitaev spin liquid to be representative of the ITO response close to the transition point, we use the exactly solvable model with a perturbative magnetic field to calculate $k$.
Later, we will discuss how our result is a more general feature of ITO.
The Kitaev model is defined through the bond-dependent Ising interactions $\H_K = \sum_{\braket{j,k}}K_{\gamma}S_j^{\gamma}S_k^{\gamma}$, where the nearest-neighbour bond $\braket{j,k}$ is of type $\gamma\in\{x,y,z\}$.
For simplicity we will focus on the isotropic limit with $|K_{\gamma}| \equiv 1$. 

We introduce a Zeeman field through $\H_Z = -h\sum_{j}\hh\cdot\gh\cdot\S_j,$ where $h = g\mu_B H$ is the effective field strength, and $\gh$ encodes the $g$-factor anisotropy.
Focusing on in-plane magnetic fields, we can neglect $g$-factor anisotropy because $g_a = g_b$. 
To retain the exact solvability of the model, the Zeeman term can be captured perturbatively in the zero-flux sector as
\begin{equation}
  \H_Z^{\mathrm{eff}} = -\frac{h_xh_yh_z}{\Delta_F^2}\sum_{(j,k,l)}S_j^{x}S_k^{y}S_l^{z},
\end{equation}
where $\Delta_F \simeq 0.065|K|$ is the two-flux gap~\cite{kitaev2006anyons,knolle2014dynamics}, and $\braket{\braket{j,l}}$ forms a second-nearest-neighbour bond.

The in-plane magnetic field direction is parametrized as $\hh = \cos(\vp)\a + \sin(\vp)\b$, and the $\c$ component of the torque and the corresponding magnetotropic coefficient $k$ are obtained by differentiating the free energy $\F$ with respect to $\vp$
\begin{equation}
  \tau_{c^*} = \t\cdot\c = \frac{\partial\F}{\partial\vp}, \qquad
  k = \frac{\partial\tau_{c^*}}{\partial\vp} = \frac{\partial^2\F}{\partial\vp^2}, 
\end{equation}
where the free energy at $T = 0$ is given by the integral $\F[\h] = -\int_{\tfrac{1}{2}\mathrm{BZ}}\d\q \sqrt{|\xi_{\q}|^2 + |\Delta_{\q}|^2}$
(over half the Brillouin zone due to reality of the MFs) involving certain well-known functions of $\q$ listed in Appendix~\ref{sec:appendixA}.
Results of this calculation are shown in Fig.~\ref{fig1}(b-c) and Fig.~\ref{fig2} as a function of $\vp$ for different values of $h$ assuming an $R\overline{3}$ space group. 

The free energy displays deep minima at the center of the ITO phases, and peaks along the $C_2$-symmetric directions where the MF mass term vanishes.
Inside the ITO, the MFs acquire a gap $\propto h^{3}$ and a non-zero Chern number $\nu = \sgn(h_xh_yh_z) = \pm 1$ for a generic field. 
Along the $\b$ axis ($[\overline{1}10]$) of $C_2$ symmetry (and equivalent directions under $C_3$) the gap vanishes, consistent with the symmetry constraints on ITO. 
Due to these extremal points, the $\c$ component of the torque vanishes along these directions, as well as along the crystallographic axes.
As we can see from the contour lines in Fig.~\ref{fig2}(b) and the slices in Fig.~\ref{fig1}(c), $\tau_{c^*}$ changes much more rapidly around the $C_2$-symmetric directions, resulting in a sawtooth-like profile.
Of the three quantities in Fig.~\ref{fig2}, the magnetotropic coefficient $k$ signals the ITO transition lines most clearly.
Midpoints of the ITO phases are marked by a broad, positive maxima in $k$ while the transition points show a sharp, \emph{negative} cusp.
While a positive value of $k$ indicates stability with respect to a rotation of the field, negative values point to a rotational instability, as a small variation in $\vp$ at these points will drive the system into one of the neighbouring $\nu = \pm 1$ phases. \par

Insight into the origin of this cusp can be obtained by examining the behaviour of the MF free energy close to the phase boundary.
The previous expression for $\F$ captures the full contribution of the MF band structure to the free energy, but the most dramatic change close to the phase boundary is the closing and re-opening of the conical dispersion at the $\bm{K}$ point.
Expanding the free MF Hamiltonian around this point, we obtain an effective continuum Hamiltonian
\begin{equation}
\H \simeq \int_{v_Fq\le\Lambda}\d\q\ \Wf_{\q}^{\dag}\left[v_F(q_x\tau_y + q_y\tau_x) + m \tau_z\right]\Wf_{\q},
\end{equation}
with $v_F = -\tfrac{\sqrt{3}}{4}K$ and
\begin{equation}
m \simeq -\frac{3\sqrt{3}}{4}\frac{h_xh_yh_z}{\Delta_F^2} = \frac{h^3}{4\sqrt{2}\Delta_F^2}\cos(3\vp),
\end{equation}
valid up to some high-energy cutoff $\Lambda$.
The dispersion is clearly $\ve_{\q} \simeq \pm\sqrt{(v_Fq)^2 + m^2}$, which we can integrate to obtain the low-energy contribution
\begin{equation}
  \F[\h] \sim \frac{2\pi}{3v_F^2}\left(|m|^{3} - (\Lambda^2 + m^2)^{3/2}\right) \sim |m|^3.
\end{equation}
From this we identify $|m|^3$ as the singular contribution to $\F$, inherited by $\tau_{c^*}$ and $k$ through derivatives with respect to $\vp$.
Close to the transition points at $\vp^* = \tfrac{\pi}{6}(2n+1)$ we have $m \propto \delta\vp = \vp-\vp^*$, and keeping only the singular contributions we find
\begin{equation}
\tau_{c^*} \sim m|m|, \qquad k \sim |m|, \qquad \partial_{\vp}k \sim \sgn(m).
\end{equation}
Therefore, the cusp singularity in $k$ is a direct consequence of the bulk MF gap closing and re-opening as a function of the field angle.
This result only depends on the assumption that the low-energy degrees of freedom are massive Dirac fermions, whose gap closes at the transition point.
We can also see that this effect is more pronounced at larger magnetic fields since MF mass on either side of the transition is larger.
Further derivatives only magnify this singularity; for instance the derivative of $k$ with respect to $\vp$ exhibits a jump discontinuity across the $C_2$-symmetric directions. \newline

\noindent{\it Summary and Open Issues} -- The observed sign change in $\kappa_{xy}$ across the $\b$-axis by \citet{yokoi2020anomalous} in \rucl\ suggests a direct transition between ITOs with opposite Chern numbers.
This is an important development, but we are currently lacking evidence of the phase transition through any thermodynamic quantities in the intermediate phase.
Here we propose the magnetotropic coefficient $k$ as a {\it sensitive} probe of a direct transition between ITOs.
Using Kitaev's exact solution with a perturbative magnetic field as a prototypical example of ITO, we have demonstrated a singular cusp in $k$ (i.e. a discontinuous jump in its derivative) at a direct transition between ITOs with opposite Chern numbers.
The origin of the cusp was shown to be directly linked to the closing of the bulk MF mass gap, and its enhancement tied to the magnitude of this gap. \par

Remarkably, extensive measurements of $k$ in \rucl\ have been amassed by ~\citet{modic2019scale} for a wide range of magnetic field strengths and orientations.
While the existing in-plane magnetotropic coefficient data in the intermediate phase is suggestive, a more detailed study around the $\b$ axis at lowest temperatures is necessary to test our proposal.
As we mentioned earlier, the torque is required to vanish along the crystallographic directions.
This presents a practical challenge, as a small signal is more susceptible to noise.
We therefore advocate for the use of resonant torque magnetometry~\cite{modic2018resonant} to directly measure the magnetotropic coefficient.
It is our hope that a comprehensive mapping of the intermediate phase boundaries with respect to the field angle can shed light on the puzzle of \rucl. 

The limitations of our result in relation to a microscopic model deserves some discussion.
Our result is limited to a direct transition between ITOs with opposite $\nu$.
The Kitaev model displays such transitions, so if a Kitaev-dominant model is appropriate for \rucl\ then a sign change of $\kappa_{xy}$ 
and our proposed singularity in $k$ across $\b$ should be observed.
Do these experimental results imply that non-Kitaev terms are negligible?
They do not necessarily imply that the underlying spin model is Kitaev dominant.
In other words, they are necessary but not sufficient; they are consequences of the Kitaev model, but do not rule out the possibility of other microscopic models with ITO. 
Note that ITO is prohibited for fields along $\b$-axis for any generic model preserving the $C_2$ symmetry. Thus the experimental signatures
discussed here are present, if ITO exists in a generic model.
While there is no microscopic model that leads to a Kitaev spin liquid under intermediate {\it in-plane} magnetic fields, if such a route exists, the model would inevitably exhibit a transition - either between two ITOs or to an intermediate phase, unless the $C_2$ symmetry is spontaneously broken.
In the absence of a microscopic understanding of the putative intermediate ITO, one cannot rule out the possibility of other significant interactions leading to an alternative mechanism, and such a microscopic model remains to be found. 

\begin{acknowledgments}
We would like to thank S. R. Julian and Y.-J. Kim for useful discussion. 
This work was supported by the Natural Sciences and Engineering Research Council of Canada and the Center for Quantum Materials at the University of Toronto.
\end{acknowledgments}

\appendix
\renewcommand\theequation{A\arabic{equation}}
\section{Exact Solution of the Kitaev Model\label{sec:appendixA}}

In this section we review the well-known solution\cite{kitaev2006anyons} of the Kitaev model
\begin{equation}
  \H_K = \sum_{\braket{j,k}}K_{\gamma}S_j^{\gamma}S_k^{\gamma},
\end{equation}
and establish the equations referenced in the main text.
Introducing four Majorana fermions $c_j^x,c_j^y,c_j^z,c_j$, the spin operators can be faithfully represented as
\begin{equation}
  S_j^{\alpha} = \frac{i}{2}c_j^{\alpha}c_j, 
\end{equation}
by restricting ourselves to the physical subspace defined through $\D_j \equiv 1$ $\forall j$ with $\D_j = \tfrac{1}{i}\s_j^x\s_j^y\s_j^z = c_j^xc_j^yc_j^zc_j$.
In this language, the spin Hamiltonian takes the form
\begin{equation}
  \H_K = \frac{i}{4}\sum_{\braket{j,k}}K_{\gamma}\hat{u}_{jk}c_jc_k,
\end{equation}
with the bond operators $\hat{u}_{jk} = -i c_j^{\gamma}c_k^{\gamma}$ that \begin{enumerate*}[label=(\roman*)] \item commute with each other, \item commute with the Hamiltonian, and \item $(\hat{u}_{jk})^2 = \id$\end{enumerate*}, furnishing an extensive set of conserved quantities.
These bond variables serve as an emergent static $\Z_2$ gauge field coupled to the $c_j$ fermions, and are related to the conserved plaquette operators in the original spin model via
\begin{equation}
  W_p = \s_j^x\s_k^y\s_l^z\s_m^x\s_n^y\s_o^z = \prod_{(j,k)\in p} \hat {u}_{jk}.
\end{equation}
The ground state lies in the flux-free sector with $W_p \equiv +1$ $\forall p$ realized with the gauge $\hat{u}_{jk} \equiv +1$ on all bonds - thus reducing the Hamiltonian to a quadratic form in the $c_j$ fermions
\begin{equation}
  \H_K = \frac{i}{4}\sum_{\braket{j,k}}K_{\gamma}c_{j,A}c_{k,B}.
\end{equation}
Here we take $\hat{u}_{j,k} = +1$ with the convention that $j$ belongs to sublattice A and $k$ belongs to sublattice B. 
Exploiting translational invariance, we may Fourier transform the Majorana fields $c_j$ via
\begin{equation}
  c_j = \sqrt{\frac{2}{N}}\sum_{\q} e^{i\q\cdot\x_j}c_{\q}, \quad c_{\q} = \frac{1}{\sqrt{2N}}\sum_j e^{-i\q\cdot\x_j}c_j,
\end{equation}
where $N$ is the number of unit cells, and the anticommutation relations $\{c_j,c_k\} = 2\delta_{j,k}$ become $\{c_{\q},c_{\p}\} = \delta_{\q,-\p}$.
The reality of the $c_j$ fermions implies $c_{\q}^{\dag} = c_{-\q}$, so the sum over $\q$ is restricted to half of the honeycomb Brillouin zone.
In momentum space the Hamiltonian in the flux-free sector becomes
\begin{equation}
  \H_K = \sum_{\q}C_{\q}^{\dag}\H_{\q}C_{\q},
\end{equation}
where $C_{\q} = (c_{-\q,A},c_{-\q,B})^T$, $C_{\q}^{\dag} = (c_{\q,A},c_{\q,B})$,
\begin{equation}
  \H_{\q} = \pmat{0 & \xi_{\q} \\ \xi_{\q}^* & 0},
\end{equation}
and $\xi_{\q} = \tfrac{i}{2}(K_x e^{-i\q\cdot\bm{a}_1} + K_ye^{-i\q\cdot\bm{a}_2} + K_z)$.
We now incorporate the effect of a Zeeman field in the zero-flux sector via
\begin{equation}
\H_Z^{\mathrm{eff}} = -\frac{h_xh_yh_z}{\Delta_F^2}\sum_{(j,k,l)}S_j^xS_k^yS_l^z.
\end{equation}
In terms of the Majorana fermions, the three-spin interaction can be written as
\begin{equation}
	S_j^xS_k^yS_l^z = \frac{i}{2^3}\hat{u}_{jk}\D_k\hat{u}_{lk}c_jc_l = \pm\frac{i}{2^3}c_jc_l,
\end{equation}
which contributes a second nearest-neighbour akin to the Haldane term.
\begin{center}
  \includegraphics[scale=0.35]{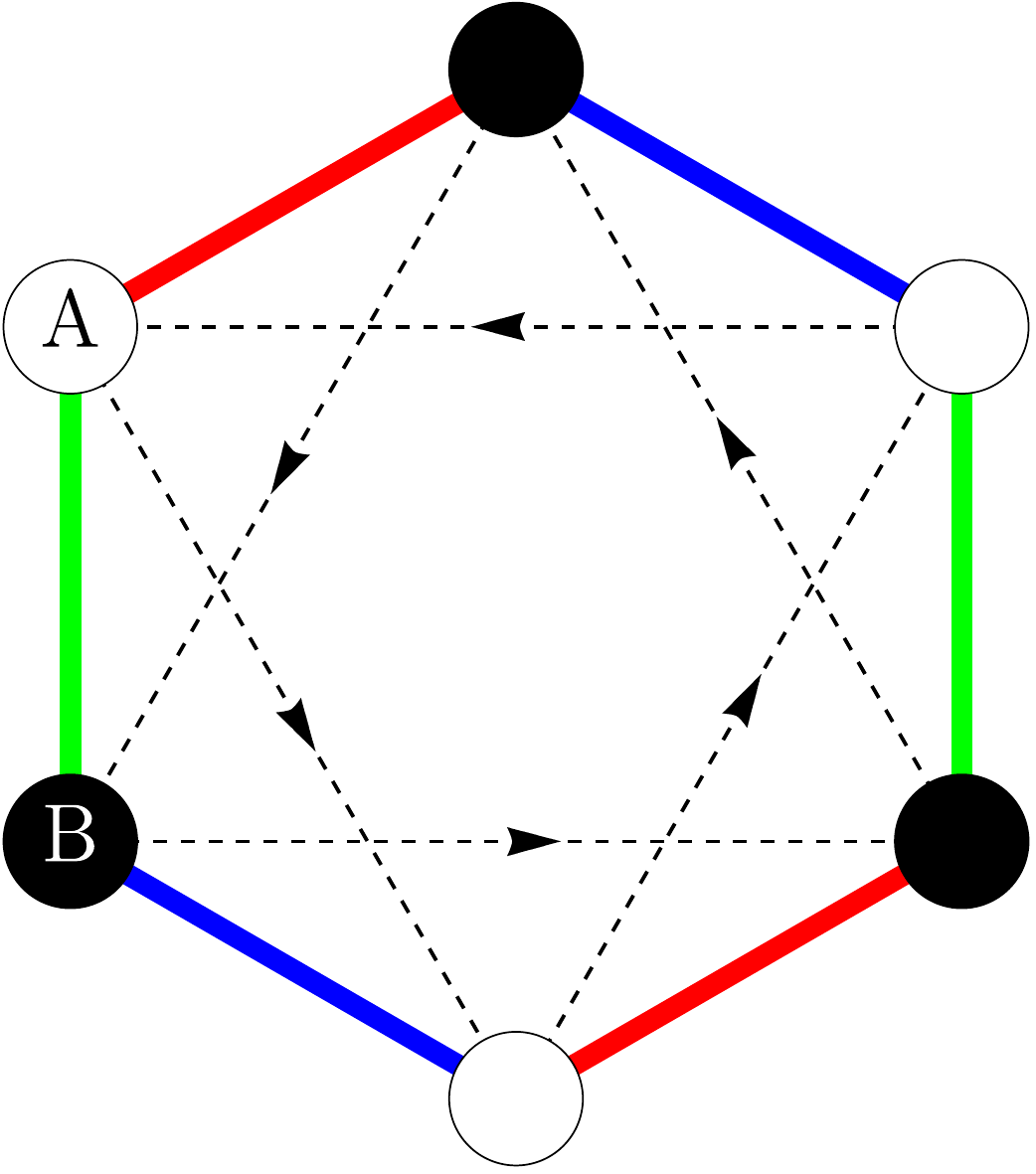}
\end{center}
Hopping along the arrow direction in the above figure contributes $+i\kappa c_jc_k$ (with opposite sign going against this direction) where the amplitude is
\begin{equation}
  \kappa = \frac{h_xh_yh_z}{2^3\Delta_F^2}.
\end{equation}
The Hamiltonian then acquires a sublattice-dependent hopping term
\begin{equation}
\H_{\q} = \pmat{\Delta_{\q} & \xi_{\q} \\ \xi_{\q}^* & -\Delta_{\q}},
\end{equation}
with
\begin{equation*}
\Delta_{\q} = 4\kappa\left[\sin(\q\cdot\bm{a}_1) - \sin(\q\cdot\bm{a}_2) - \sin(\q\cdot(\bm{a}_1-\bm{a}_2))\right].
\end{equation*}
We now consider $K_x = K_y \neq K_z$, consistent with $C2/m$ symmetry, but sufficiently close to the isotropic limit ($|K_x| \ge \tfrac{1}{2}|K_z|$) so that the Majorana spectrum remains gapless for $h = 0$.
The position of the node at $h = 0$ is easily found to be at $\q_* = (1-q,q)$ in the reciprocal lattice basis, where
\begin{equation}
  q = \frac{1}{2\pi}\arccos\left(-\frac{1}{2}\frac{K_z}{K_x}\right).
\end{equation}
In the isotropic limit this reduces to $\q_* = (\tfrac{2}{3},\tfrac{1}{3})$ - i.e. the $\bm{K}$ point of the honeycomb Brillouin zone.
Expanding the Majorana Hamiltonian about $\q_*$ yields the continuum Hamiltonian 
\begin{equation}
\H \simeq \int_{v_Fq\le\Lambda}\d\q\ \Wf_{\q}^{\dag}\left[v_F(\eta q_x\tau_y + q_y\tau_x) + m \tau_z\right]\Wf_{\q},
\end{equation}
valid up to an energy cutoff $\Lambda$, where
\begin{equation*}
  v_F = -\frac{\sqrt{3}}{4}K_z, \quad \eta = \frac{1}{\sqrt{3}}\sqrt{4\left(\frac{K_x}{K_z}\right)^2 - 1},
\end{equation*}
and
\begin{equation*}
m \simeq -6\sqrt{3}\kappa = -\frac{3\sqrt{3}}{4}\frac{h_xh_yh_z}{\Delta_F^2}.
\end{equation*}
Note that there is a slight anisotropy in the velocity captured by the $\eta$ parameter.
However, this effect is small (with $K_{x/y}/K_z \simeq 0.95$, $\eta \simeq 0.93$) and does not affect the qualitative features of our result, so we neglect it for simplicity.
The magnetic field term $\Delta_{\q}$ also introduces a linear term in $\q$ away from the isotropic limit, but it is safe to ignore because $\kappa$ is assumed to be small.

\bibliography{references}\label{sec:bib}

\end{document}